\documentclass[sigconf,10pt]{acmart}

\usepackage[ruled,linesnumbered,vlined]{algorithm2e}
\usepackage{subcaption} 
\usepackage{hyperref}
\usepackage{url}
\newtheorem{fact}{Fact}

\AtBeginDocument{%
  \providecommand\BibTeX{{%
    \normalfont B\kern-0.5em{\scshape i\kern-0.25em b}\kern-0.8em\TeX}}}

\copyrightyear{2024}
\acmYear{2024}
\setcopyright{acmlicensed}\acmConference[ICS '24]{Proceedings of the 38th ACM International Conference on Supercomputing}{June 4--7, 2024}{Kyoto, Japan}
\acmBooktitle{Proceedings of the 38th ACM International Conference on Supercomputing (ICS '24), June 4--7, 2024, Kyoto, Japan}
\acmDOI{10.1145/3650200.3656600}
\acmISBN{979-8-4007-0610-3/24/06}

\begin{document}

\title{DAWN: Matrix Operation-Optimized Algorithm for Shortest Paths Problem on Unweighted Graphs}






\author{
	Yelai Feng\textsuperscript{\textrm{1, 2}},
	Huaixi Wang\textsuperscript{\textrm{1, *}},
	Yining Zhu\textsuperscript{\textrm{3}},
	Xiandong Liu\textsuperscript{\textrm{4}},
	Hongyi Lu\textsuperscript{\textrm{2}},
	Qing Liu\textsuperscript{\textrm{5}},
}

\email{{fengyelai,wanghuaixi}@nudt.edu.cn}

\affiliation{
	\textsuperscript{\textrm{ 1}} College of Electronic Engineering, National University of Defense Technology, Hefei, China\\
	\textsuperscript{\textrm{ 2}}College of Computer Science and Technology, National University of Defense Technology, Changsha, China\\
	\textsuperscript{\textrm{ 3}}Ningbo Institute of Technolog, Zhejiang University, Ningbo, China\\
	\textsuperscript{\textrm{ 4}}PerfXLab Technologies Co.,Ltd, Beijing, China\\
	\textsuperscript{\textrm{ 5}}College of Electrical and Computer Engineering, Technical University of Munich, Munich, Germany\\
	\country{}
}

\renewcommand{\shortauthors}{Yelai Feng and Huaixi Wang, et al.}

\begin{abstract}
	The shortest paths problem is a fundamental challenge in graph theory, with a broad range of potential applications. The algorithms based on matrix multiplication exhibits excellent parallelism and scalability, but is constrained by high memory consumption and algorithmic complexity. Traditional shortest paths algorithms are limited by priority queues, such as BFS and Dijkstra algorithm, making the improvement of their parallelism a focal issue. We propose a matrix operation-optimized algorithm, which offers improved parallelism, reduced time complexity, and lower memory consumption. The novel algorithm requires $O(E_{wcc}(i))$ and $O(S_{wcc} \cdot E_{wcc})$ times for single-source and all-pairs shortest paths problems, respectively, where $S_{wcc}$ and $E_{wcc}$ denote the number of nodes and edges included in the largest weakly connected component in graph. To evaluate the effectiveness of the novel algorithm, we tested it using graphs from SuiteSparse Matrix Collection and Gunrock benchmark dataset. Our algorithm outperformed the BFS implementations from Gunrock and GAP (the previous state-of-the-art solution), achieving an average speedup of 3.769$\times$ and 9.410$\times$, respectively.
\end{abstract}




\maketitle

\section{Introduction}
The shortest paths problem, a fundamental problem in graph theory and network science, has garnered interest from researchers across various disciplines such as transportation planning, computer science, network science, and applied mathematics \cite{dreyfus1969appraisal,bertsekas1998network,tarjan1983data,waissi1994network,bertsekas1991analysis}. As the scale of the graph increases, serial algorithms struggle to adapt to changes, and prompting researchers to explore parallel computing as a solution to the shortest paths problem.

The state-of-the-art solution for SSSP (Single-Source Shortest Paths) problem is the BFS algorithm on unweighted graph and $\Delta$-stepping Dijkstra's algorithm on weighted graph \cite{meyer2003delta}, which have garnered significant attention \cite{chan2012all,davidson2014work,busato2015efficient,surve2017parallel,wang2019sep}. Currently, there are several solutions available in the industry for rapidly computing the APSP (All-Pairs Shortest Paths) problem on large-scale clusters \cite{10.1145/3431379.3460651,sao2020supernodal}.

Timothy M et al. proposed two novel APSP algorithms based on the BFS algorithm, which require $O(\frac{mn}{\log n})$ ($m \gg n \log^2 n$) and $O(\frac{mn \log \log n}{ \log n} + \frac{n^2 \log^2 \log n}{ \log n})$ times for all graphs \cite{chan2012all}, where $m$ and $n$ respectively represents the number of nodes and edges in graph. The APSP algorithms based on matrix multiplication, such as Seidel's algorithm \cite{seidel1995all}, Galil and Margalit's algorithm \cite{GALIL1997103}, reduces the time complexity by divide-and-conquer strategy, these approaches require significant memory resources for maintaining intermediate matrices.

We introduce a novel algorithm named DAWN (Distance Assessment algorithm With matrix operations on Networks), which is based on matrix operation-optimizing. DAWN requires $O(m)$ space and $O(E_{wcc}(i))$ time on unweighted graphs for SSSP tasks. It is also capable of processing APSP tasks and requires $O(S_{wcc} \cdot E_{wcc})$ time. Here, $S_{wcc}$ and $E_{wcc}$ denote the number of nodes and edges included in the largest WCC (Weakly Connected Component) in the graphs, and $i$ is the source node of the SSSP task.

The main contributions of this work are as follows:

\begin{enumerate}
	\item   We propose a matrix operation-optimized algorithm, which requires $O(m)$ space and $O(E_{wcc}(i))$ times on the unweighted graphs for SSSP problem, respectively. In contrast to the prevalent optimization of state-of-the-art BFS implementation, which commonly rely on priority queues, our approach leverages matrix operations to endow DAWN with enhanced parallelism.
	\item   We propose BOVM (Boolean Vector-Matrix Operation) method, which make DAWN to require $O(\frac{\epsilon(i)}{2}m)$ time for SSSP tasks on unweighted graphs, where $\epsilon(i)$ is the eccentricity of node $i$. Further, we propose an SOVM (Sparse Optimized Boolean Vector-Matrix Operation) method to significantly improve the performance of DAWN on sparse graphs,  reducing the time requirements to $O(E_{wcc}(i))$ for SSSP tasks and $O(S_{wcc} \cdot E_{wcc})$ for APSP tasks.
	\item   DAWN achieves superior performance compared to Gunrock while utilizing fewer GPU memory resources. It successfully completes the SSSP task on graphs with 214 million nodes and 936 million edges using an RTX 3080TI, a task unattainable by Gunrock. Prior to DAWN, algorithms based on matrix multiplication used a divide-and-conquer strategy, such as Seidel's algorithm \cite{seidel1995all}, which generated numerous intermediate matrices and required excessive memory.
\end{enumerate}

In Section 2, we present an overview of the typical shortest paths algorithms. In Section 3, we describe the design of the DAWN and propose optimization methods to make it more widely applicable to various graphs. In Section 4, we conducted comparative experiments of several implementations across various platforms, which demonstrates the high efficiency of DAWN. In Section 5, we summarize the work in this paper and outline future research directions. Table \ref{tab:notations} lists the notations used throughout the paper.

\begin{table}[htpb]
	\begin{center}
		\begin{minipage}{\linewidth}
			\caption{Definition of notations}\label{tab:notations}%
			\begin{tabular*}{\textwidth}{@{\extracolsep{\fill}}cl@{\extracolsep{\fill}}}
				\hline
				Notation & Definition	  \\
				\hline
				$A$& Adjacency matrix of unweighted graphs \\
				$E_{wcc}$& Maximum edge count of the largest WCC\\
				$S_{wcc}$& Maximum node count of the largest WCC\\
				$E_{wcc}(i)$& Edge count of the largest WCC of node $i$\\
				$S_{wcc}(i)$& Node count of the largest WCC of node $i$\\
				$\epsilon(i)$& Eccentricity of node $i$\\
				$\epsilon_{max}$& Maximal eccentricity of the graph\\
				$n,m$& Number of nodes and edges in graph\\
				$p$& Average connected probability\\
				$CSR$& Compressed Sparse Row format\\
				$CSC$& Compressed Sparse Column format\\
				\hline
			\end{tabular*}
		\end{minipage}
	\end{center}
\end{table}

\section{Related Works}
\label{relatedworks}
The shortest paths problem is a classic problem in graph theory and network science. In this section, we introduce the typical algorithms for solving the SSSP and APSP problems. In addition, we will introduce several APSP algorithms based on matrix multiplication.

\subsection{SSSP algorithm}
Dijkstra's algorithm is a common SSSP algorithm \cite{dijkstra1959note}, and the main optimization methods are priority queue of binary heap and Fibonacci heap \cite{johnson1977efficient,raman1997recent,thorup2000ram,fredman1987fibonacci,moret1992empirical,chen2007priority}.

Meyer et al. proposed an optimized Dijkstra's algorithm, which is a parallel version for a large class of graphs \cite{meyer2003delta}. The best parallel version of the  $\Delta$-stepping Dijkstra's algorithm takes $O(D\cdot L \cdot \log n + \log^2 n)$ time and $O(n + m + D\cdot L \cdot \log n)$ work on average, where $L$ denotes the maximum shortest paths weight from the source node ($s$) to any node reachable from $s$, and $D$ represents the maximum node degree \cite{meyer2003delta}.



\subsection{BFS Algorithm}
Scott Beamer et al. proposed a hybrid BFS algorithm that combines a conventional top-down approach with a novel bottom-up approach \cite{beamer2012direction}. In the top-down approach, nodes within the active frontier seek unvisited child nodes, whereas in the bottom-up approach, unvisited nodes seek parents within the active frontier. Scott Beamer et al. further optimized the performance of direction-optimized BFS in various application scenarios \cite{beamer2013distributed,bulucc2017distributed}, ultimately integrating it into GAP (Graph Algorithm Platform benchmark Suite). GAP is a portable high-performance baseline which includes representative implementations of state-of-the-art performance, and is intended to help graph processing research by standardizing evaluations\cite{beamer2015gap}.

Julian Shun and Laxman Dhulipala et al. inspired by the direction-optimized BFS algorithm, achieve close to the same efficiency (time and space) as the optimized BFS of Beamer et. al. and using a much simpler implementation code \cite{shun2013ligra,shun2015smaller}. Further, they promoted widespread applications of this optimization approach and constructed a high-performance computing framework called GBBS (Graph Based Benchmark Suite) which based on the Ligra/Ligra+/Julienne graph processing frameworks \cite{shun2016parallel,dhulipala2017julienne,shun2020practical,dhulipala2018theoretically}, including betweenness centrality, graph radius estimation, graph connectivity, PageRank and single-source shortest paths etc..

\subsection{Matrix algorithm}
The shortest paths algorithm based on matrix squaring multiplication has received extensive attention \cite{farbey1967cascade,takaoka1998subcubic,mulmuley2000lower}. The matrix squaring multiplication algorithm reduces the number of matrix multiplications from $n$ to $\log n$, but it requires storing many intermediate matrices. When the graph is large, the algorithm needs to consume a significant amount of space.

Seidel's algorithm is the APSP algorithm based on matrix multiplication \cite{seidel1995all}, which is suitable for unweighted undirected graphs and time complexity is $O(\log n \cdot n^{\omega})$. The Seidel's algorithm requires numerous memory to maintain intermediate matrices, due to reduce the time complexity of GEMM (General Matrix-Matrix multiplication) via divide-and-conquer strategy \cite{strassen1969gaussian, 10.1145/28395.28396, williams2012multiplying}.

Arlazarov et al. proposed the APSP algorithm based on the boolean matrix multiplication on the unweighted graphs \cite{arlazarov1970economical}, which has alleviated the issue of excessive memory requirements associated with matrix multiplication-based APSP algorithms to a limited extent.


\section{Methods}

In this section, we will introduce the DAWN and the technical details of BOVM and SOVM. In Section \ref{Sec:Principle}, we will illustrate the principle of DAWN. We will then discuss the optimization of boolean vector-matrix operation (BOVM) that enables DAWN's efficiency in Section \ref{Sec:BOVM}. Furthermore, we will expand on BOVM and introduce its extension to sparse matrices, known as sparse optimized boolean vector-matrix operation (SOVM), in Section \ref{Sec:SOVM}. In Section \ref{Sec:Difference}, we use an example to demonstrate difference between BFS and DAWN.

\subsection{Principle} \label{Sec:Principle}
DAWN relies on the result of matrix multiplication to assist in determining which edge visits can be skipped. However, matrix multiplication is a costly operation, requiring $O(n^3)$ time and $O(n^2)$ space.

Our main contribution is the simplification of matrix multiplication, which does not mean that we can compute matrix multiplication faster, but rather that we focus on only a portion of the results of matrix multiplication for the shortest path problem. Specifically, our new approach is able to determine which rows and columns of the matrix multiplication result have an impact on the shortest path problem.

\begin{figure}[htbp]
	\centering
	\includegraphics[width=\linewidth]{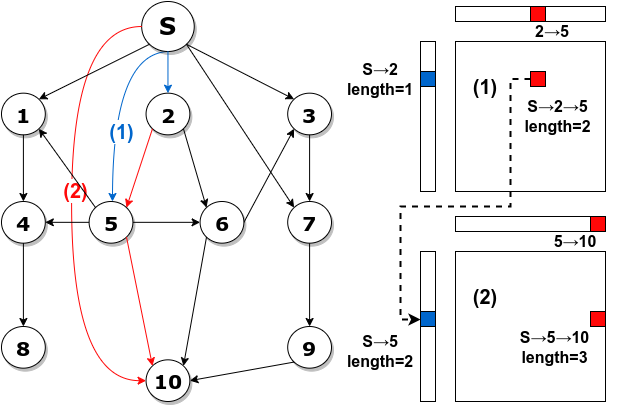}
	\caption{Example for the Shortest Paths Discovery in the Boolean Matrix}
	\label{fig:matrix_example}
\end{figure}

Figure \ref{fig:matrix_example} illustrates the correspondence between Boolean matrix operations and shortest path discovery in a graph. The blue markers indicate the result vector, while the red markers indicate a particular row of the adjacency matrix.

\begin{lemma}\cite{harary1965structural}
	In the matrix $A^{k} = \big(a_{i,j}^{(k)}\big)_{n\times n}$, the element $a_{i,j}^{(k)}$ represents the number of paths with length $k$ from $v_i$ to $v_j$.
	\label{path}
\end{lemma}

\begin{theorem}
	In unweighted graphs, the length of the shortest path from $v_i$ to $v_j$ is $k_{min}$, if and only if $a_{i,j}^{(k_{min})}\not =0 \land i\not=j \land \sum_{k=1}^{k_{min}-1}a_{i,j}^{(k)}=0$.
	\label{sp}
\end{theorem}

\begin{fact}
	In unweighted graphs, any shortest paths of length $k$ can be expressed as the connection of two shortest paths with lengths $k-1$ and $1$, where $k\geq 2$.\label{fact1}
\end{fact}

Obviously, we can obtain an expression for all paths,
\begin{equation}
	\begin{split}
		P(i,j)_{min} &=\sum_{k=1}^{\epsilon(i)}a_{i,j}^{(k)}=\sum_{k=1}^{k_{min}-1}a_{i,j}^{(k)}+a_{i,j}^{(k_{min})}+\sum_{k=k_{min}+1}^{\epsilon(i)}a_{i,j}^{(k)},
	\end{split}
\end{equation}
where $1\leq k_{min} \leq \epsilon(i)$. It is evident that the sum of array is minimized when the first non-zero value of $a_{i,j}^{(k)}$ is encountered.

Theorem \ref{sp} and Fact \ref{fact1} state the sufficient condition for breaking the loop and ending it:
\begin{enumerate}
	\label{condition}
	\item DAWN already found the shortest paths between all pairs of nodes in the graph;
	\item The distance vector does not change when a loop ends,  which means no new paths were found in the loop.
\end{enumerate}

\subsection{BOVM}\label{Sec:BOVM}

We describe the vector-vector multiplication as follows,
\begin{equation}
	a_{i,j}^{(2)}=a_{i,1}a_{1,j}+ a_{i,2}a_{2,j} \cdots + a_{i,n}a_{n,j},\label{a2}
\end{equation}
which denotes the collection of path combinations from $i$ to $j$ through any node. Thus, it is unnecessary to consider all possible combinations when determining the presence of a path from i to j; only cases where$a_{i,l}>0 \wedge a_{l,j}>0$ can affect the value of $a_{i,j}^{(2)}$. If $a_{i,j}^{(2)}$ represents any value greater than 0, it signifies the existence of a shortest path from node $i$ to $j$. Consequently, we can simplify this formula by utilizing a Boolean data type.

We converted the Formula \ref{a2} to a Boolean-type as follows,
\begin{equation}\label{ab}
	a_{i,j}^{(2)}= \sum_{l=0}^{n-1}\alpha[l] \wedge \beta[l],
\end{equation}
which requires $O(n)$ time. Since only the non-zero elements in $\alpha$ and $\beta$ affect the multiplication result, we can compress the vectors by retaining only their non-zero elements, and get Formula \ref{index} as follows,
\begin{equation}\label{index}
	a_{i,j}^{(2)}= \sum_{l=0}^{len_{\gamma}-1}\alpha[\gamma[l]],
\end{equation}
where $len_{\gamma}$ represents the length of $\gamma$ and $\gamma$ is a compressed version of $\beta$, containing only the indices of non-zero elements.

\begin{algorithm}[htbp]
	\caption{Boolean Vector-Matrix Operation}
	\label{BOVM}
	\SetAlgoLined
	\LinesNumbered
	\KwIn{ CSC, $\alpha$, $\beta$, distance, step, is\_converged}
	\KwOut{distance, is\_converged}
	\While{step < n} {
		step $\Leftarrow$ step + 1 \;
		\For{i $\in$ [0,n-1] \& $\alpha$[i] = false}
		{
			start $\Leftarrow$ CSC.columns\_ptr[i] \;
			end $\Leftarrow$ CSC.columns\_ptr[i + 1] \;
			\While {(j $\in$ [start, end-1]) \& ($\alpha$[CSC.row[j]] = true) \& (CSC.row[k] != i)}
			{
				$\beta$[i] $\Leftarrow$ true \;
				distance[i] $\Leftarrow$ step \;
				\If{is\_converged = true}{
					is\_converged $\Leftarrow$ false\;
				}
				break\;
			}
			\tcp{The vector $\alpha$ and $\beta$ are mandatory and cannot be replaced by search vector $distance$.}
			$\alpha \Leftarrow merge(\alpha,\beta)$ \;
			$ \beta.swap(NULL) $\;
		}
		\If {is\_converged = true} {
			break \;
		}
	}
	\Return{ distance}\;
\end{algorithm}

In Formula \ref{index}, if the value of $\alpha[\gamma[l]]$ is 1, the result $a_{i,j}^{(2)}$ will be 1. Once the first element of $true$ is obtained, the sum will always yield a value of $true$ and indicates that the path exists from $i$ to $j$. Hence, we let the loop end at the time. We can ensure that the path we first discover is the shortest path by Theorem \ref{sp}, and skip the computation of any paths ended with $j$ in next operations.

We can extend this vector operation to the CSC format matrix to assist DAWN in reducing neighbor node access and discovering the shortest path. We get the BOVM as the Algorithm \ref{BOVM}, where $\alpha$ and $\beta$ are the dense vector, $step$ represents the steps of the shortest paths in the iteration. Line 4 of Algorithm \ref{BOVM} implements the Formula \ref{index}, while lines 7-8 implement the stopping criterion provided by Fact \ref{fact1}.

Next, we discuss the time complexity of DAWN based on the BOVM, and the formula as follows,
\begin{align}
	T(n) & =\sum_{x=1}^{\epsilon(i)-1}\big[\sum_{i=0}^{n-len[\alpha_{x}]}D_{exit}(i,x)\big],                      \\
	T(n) & <\sum_{x=1}^{\epsilon(i)-1}\big[1- \frac{1-p}{\epsilon(i)-1}x \big]\cdot m <\frac{1+p}{2}\epsilon(i)m.
\end{align}
where $D_{exit}(i,x)$ represents the index value of element in $A[i]$ when DAWN exits the loop $x$, $len[\alpha_{x}]$ is the length of $\alpha$ in the loop $x$ (Refer to Algorithm \ref{BOVM} Line 1 to 5), $p$ represents the average connection probability of graph.

DAWN requires $O(\frac{1+p}{2}\epsilon_{max}\cdot nm)$ and $O(\frac{1+p}{2}\epsilon(i)m)$ time for APSP and SSSP problem, respectively.

\subsection{Sparse Optimized Operation}\label{Sec:SOVM}
The performance of BOVM on sparse graphs, especially those with large diameters, is often limited by the expensive cost of vector-matrix multiplication, making it difficult to outperform BFS. Reducing the number of vector multiplications has become a critical issue in enabling DAWN to be widely used.

We propose the method of SOVM to optimized DAWN on the sparse graphs, which combines graph traversal algorithms with vector-matrix multiplication, limiting the operation to nodes and their neighboring nodes. Specifically, we first obtain the set of neighboring nodes, exclude nodes that have already appeared in the result vector, then calculate the vector multiplication values of these nodes, obtaining paths of length step with target nodes in the neighboring nodes set, and finally update the shortest paths in the result vector.

Although the process is complex, we can simplify it by utilizing the properties of Boolean matrix operations. It is important to note that SOVM operates on CSR matrices, while BOVM operates on CSC matrices. The boolean vector-matrix multiplication is as follows,
\begin{align}
	\gamma[i] = \sum_{l=0}^{n}\alpha[l] \wedge A[i][l].
\end{align}
If we use matrix $A = \{\beta_0,\beta_1, \cdots \beta_{n-2}, \beta_{n-1}\}$, we can simplify the boolean vector-matrix multiplication as follows,
\begin{align}\label{bovm}
	\gamma = \prod_{k=0}^{len_{\beta'}-1}\beta_{\beta'[k]},
\end{align}
where $\beta'$ is the compress version of $\beta$. Formula \ref{bovm} indicates that the BOVM can be achieved by computing multiple inner products of vectors in succession.

If we transpose the matrix A to a CSR matrix $A\_CSR = \{\alpha_0,\alpha_1, \cdots \alpha_{n-2}, \alpha_{n-1}\}$, Formula \ref{bovm} can be simplified as follows,
\begin{align}\label{sovm}
	\beta = \bigcup_{k=0}^{len_{\beta'}-1} \alpha_{\beta'[k]},
\end{align}
and it means that we can use $len_{\beta'}$ times of array merges to replace boolean vector-matrix multiplication in the SSSP tasks. We get the optimized method as the Algorithm \ref{SOVM}.

\begin{algorithm}[htbp]
	\caption{Sparse Optimized Boolean Vector-Matrix Operation}
	\label{SOVM}
	\SetAlgoLined
	\LinesNumbered
	\KwIn{ CSR, $\alpha$, $\beta$, step, distance}
	\KwOut{ distance}
	\While{step < n} {
		step $\Leftarrow$ step + 1 \;
		\While{i $\in$ [0, n-1] \& ($\alpha$[i] = true) }{
			start $\Leftarrow$ CSR.row\_ptr[i]\;
			end $\Leftarrow$ CSR.row\_ptr[i + 1]\;
			\While {(j $\in$ [start, end-1]) \& (distance[CSR.col[j]] = 0)}
			{
				$\beta$[CSR.col[j]]$\Leftarrow$  true\;
				distance[CSR.col[j]] $\Leftarrow$  step\;
				\If{is\_converged = true}{
					is\_converged $\Leftarrow$ false\;
				}
			}
		}
		$\alpha$.swap(NULL)\; 	\tcp{Reset the vector by swap}
		\If {is\_converged = true} {
			break \;
		}
	}
	\Return{ distance}\;
\end{algorithm}

Algorithm \ref{SOVM} utilizes a simpler method to merge vectors, and is particularly interested in the newly added elements of $\beta$ after merging these arrays. We aim to skip any duplicate elements since these shortest paths have already been discovered. We only visit the edges and update the shortest path when the element is missing in the $\beta$ array.

Specifically, SOVM starts from the set of neighbor nodes, skips all nodes that have already appeared in the result vector (line 1), finds the target nodes in neighboring nodes set that have not yet appeared in the result vector (line 4), and then updates their shortest paths. Formula \ref{sovm} provides theoretical support for such operations, and SOVM can automatically exclude the cycles without additional judgment.

Hence, we get the time complexity of DAWN based on SOVM to solve SSSP task of node $i$ is as follows,
\begin{align}
	T(n) = \sum_{j=0}^{S_{wcc}(i)}d^+(j) = E_{wcc}(i),
\end{align}
where $d^+(j)$ is the out-degree of node $j$. $S_{wcc}(i)$ and $E_{wcc}(i)$ denotes the number of nodes and edges included in the largest WCC (Weakly Connected Component) to which node $i$ belongs. The time complexity of DAWN for APSP is determined by the largest WCC in the graph,
\begin{align}
	T(n) & = \sum_{i=0}^{n-1}E_{wcc}(i)= S_{wcc} \cdot E_{wcc} + \sum_{i=0}^{n-1-S_{wcc}}E_{wcc}(i), \\
	T(n) & < 2S_{wcc} \cdot E_{wcc},
\end{align}
where $S_{wcc}$ and $E_{wcc}$ denote the number of nodes and edges included in the largest WCC in graph. The time complexity of DAWN based on the SOVM is $O(S_{wcc} \cdot E_{wcc})$ for APSP tasks.

In summary, DAWN based on SOVM achieves better time complexity, requiring $O(S_{wcc} \cdot E_{wcc})$ and $O(E_{wcc}(i))$ time for APSP and SSSP tasks on the unweighted graphs, compared to BFS which requires $O(nm)$ and $O(m)$, respectively. It is important to note that this complexity improvement only occurs in non-connected graphs, whereas in connected graphs, DAWN and BSF both require $O(nm)$ and $O(m)$ time for APSP and SSSP tasks.

\subsection{Memory}
In this section, we elaborate on how DAWN achieves reduced memory usage compared to BFS. Typically, the memory requirements of the BFS algorithm can be divided into three components: CSR matrix, the distance vector, and the priority queue. This implies that BFS cannot operate with less than $4m+8n$ bytes of memory.

DAWN's memory requirements also consist of three components: the CSR matrix, the distance vector, and two boolean arrays. The two boolean arrays are utilized to store the paths updated in the previous and current iterations (details in Algorithm \ref{SOVM}). As DAWN is a backward BFS algorithm, we can maintain a boolean array on the GPU instead of distance vector, with path length updates occurring in memory. The GPU memory is byte-addressable, and even boolean variables are allocated a byte of space.

Therefore, DAWN necessitates a minimum of $4m+3n$ bytes of memory. We can get the formula as follows,
\begin{equation}
	\eta = \frac{4m+3n}{4m+8n} = \frac{4D+3}{4D+8},
\end{equation}
where $D$ represents the average degree	of the graph.

For instance, when considering the theoretical minimum memory usage, DAWN requires only 91.58\% of the memory used by Gunrock on the graph \textit{uk-2005}. While the theoretical difference is approximately 8.4\%, in experiments conducted under constrained GPU memory conditions, DAWN can solve the BFS task on \textit{uk-2005}, whereas Gunrock fails to allocate sufficient GPU memory. It is noteworthy that as the sparsity of the graph increases, the memory advantage of DAWN becomes more pronounced.

\subsection{Difference} \label{Sec:Difference}
To further examine the differences between DAWN and BFS, we present the technical details used in these two algorithms. Algorithm \ref{BFS} describes the general BFS algorithm, and the benchmark implementations from GAP and Gunrock both employed more sophisticated optimization techniques in the experiment, where $pq$ represents the priority queue and $source$ is the source node of the SSSP task.

\begin{algorithm}[htbp]
	\caption{General BFS}
	\label{BFS}
	\SetAlgoLined
	\LinesNumbered
	\KwIn{ CSR, pq, source, distance}
	\KwOut{ distance}
	pq.push(\{source,distance\}) \;
	\While{!pq.empty()}{
		i $\Leftarrow$ pq.top; pq.pop()\;
		start $\Leftarrow$ CSR.row\_ptr[i]\;
		end $\Leftarrow$ CSR.row\_ptr[i + 1]\;
		\While {j $\in$ [start, end-1]}{
			index $\Leftarrow$ CSR.col[j]\;
			\If{distance[index] = 0}{
				new\_dist $\Leftarrow$ distance + 1\;
				distance[index] $\Leftarrow$ new\_dist\;
				pq.push(\{index,new\_dist\}) \;
			}
		}
	}
	\Return{distance}\;
\end{algorithm}

For Line 15 of Algorithm \ref{SOVM}, if no new shortest paths are found in this loop, then exit, according to Fact \ref{fact1}. We utilize \textit{step} to mark the current node being visited as a neighbor node of the source node at the layer $step$ in DAWN. Lines 4-6 of Algorithm \ref{SOVM} indicate that some edge visitations can be skipped. This means that the nodes that have already been visited in the previous layer do not need to be visited again, as the shortest path has already been determined.  The theorem supporting this decision is referred to as Theorem \ref{sp}, which states that the first discovered path from the source node to a reachable node is the shortest path. The processing steps of DAWN is as follows,
\begin{enumerate}
	\item Firstly, DAWN reads the input vector and identifies the single-step reachable nodes from the source node $s$,
	\item Then, searching for the single-step reachable nodes from the updated nodes which updating in the previous step, while skipping nodes that have already been discovered to have a path from the source node $s$ and reachable nodes with an out-degree of 0,
	\item Next, DAWN repeats the second step until output vector stabilizes.
	\item Finally, exit loop and output the result vector.
\end{enumerate}

On the other hand, in BFS, the operations of accessing nodes and edges, and checking whether the path needs to be updated are necessary for every node and edge, refer to Line 6-10 in Algorithm \ref{BFS}.

\begin{figure}[htbp]
	\centering
	\includegraphics[width=\linewidth]{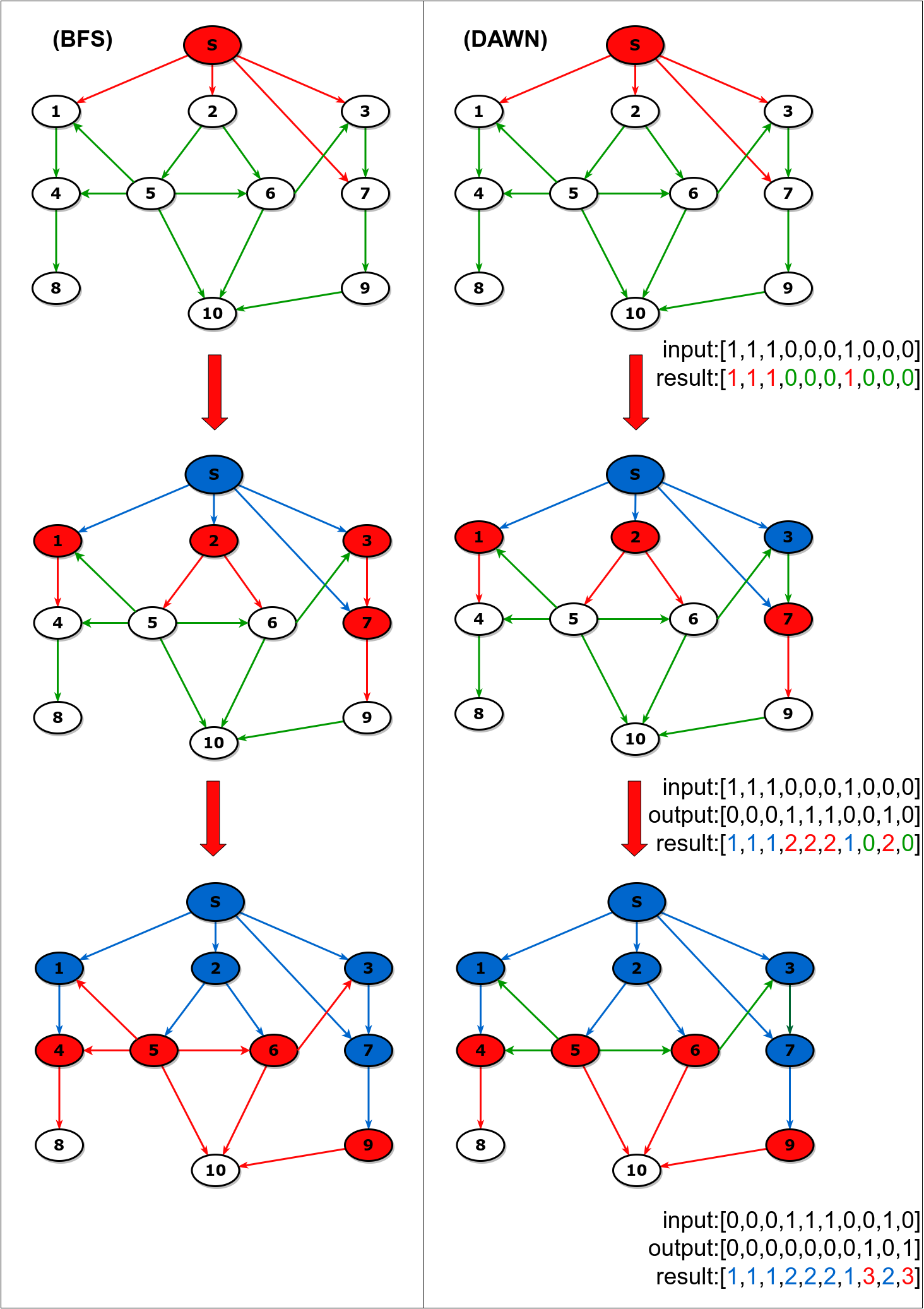}
	\caption{Example for the BFS and DAWN Processing, Left is the BFS and Right is the DAWN}
	\label{fig:example}
\end{figure}

We illustrate the difference between the BFS algorithm and DAWN through a example in Figure \ref{fig:example}. The red color indicates the nodes and edges that are visited in the current step, the blue nodes and edges represent that have already been visited, and the green edges represent that have not yet been visited. The $input$ represents the paths updated in the previous iteration, while the $output$ indicates the paths updated in the current iteration. The $result$ signifies the outcome of the algorithm, specifically denoting the shortest path lengths from the source node to other nodes in the graph.

In the third step of Figure \ref{fig:example}, node $5$ has four outbound edges to node $1,4,6,10$. BFS must traverse these edges, and then BFS would note that the destination vertex of these edges had already been visited and the destination vertex would not be in the output frontier. However, DAWN does not traverse these edges. The compressed vector for node 5 is $\{1,2,3,6,10\}$, and $result$ is $\{1,1,1,0,0,0,1,0,0,0\}$. The values of $result[1]$, $result[2]$, $result[3]$, and $result[6]$ are all 1, indicating that DAWN will skip these edges, because the shortest paths of them were already found.

In the SSSP task for node $s$, the BFS algorithm visited a total of 10 nodes and 17 edges, while DAWN visited only 8 nodes and 12 edges, resulting in 2 fewer nodes and 5 fewer edges being visited by DAWN.

Overall, the fundamental difference between DAWN and BFS lies in whether the algorithm relies on a priority queue to prevent revisiting nodes and edges.

\section{Results}
In this section, we will outline the experimental setup and present initial experimental data. Following this, we proceed to show the performance of DAWN with regard to scalability and its effect in accelerating the SSSP task.

\subsection{Experiment Introduction}\label{exp}
In our experimental trials, we leverage a set of 66 general graphs sourced from the SuiteSparse Matrix Collection and the Gunrock benchmark datasets \cite{Davis2011matrix,10.1145/2851141.2851145}.

In scenarios where a node is not part of the largest WCC of the graph, but instead resides in a smaller connected component, DAWN has the potential to accomplish the task in constant time, while BFS requires the construction of a priority queue.

Therefore, we have established a randomly generated set comprising 500 nodes, where each node executes the SSSP task 64 times. All computations are conducted within this node-set. It is noteworthy that these nodes are not exclusively part of the largest connected component, and our dataset includes non-connected graphs. We underscore our focus on evaluating the performance of the BFS algorithm across diverse graph types, including connected and non-connected graphs, as well as both generated and real-world graphs.

Performing the task consecutively serves to minimize the influence of external factors on experimental results, such as interference from background processes. We adopted the arithmetic mean as the anticipated value and subjected the sample distribution to a t-distribution test. After eliminating samples that deviated from the assumptions of the t-distribution, and computing the mean of the remaining samples, we get the final result.

\begin{table}[htbp]
	\begin{center}
		\begin{minipage}{\linewidth}
			\caption{Parameters of the Test Machine}\label{TestMachine}%
			\begin{tabular*}{\textwidth}{@{\extracolsep{\fill}}ccc@{\extracolsep{\fill}}}
				\toprule
				Hardware & Machine1&Machine2\\
				\midrule
				CPU	& Intel Core i5-13600KF& AMD EPYC Milan 7T83\\
				RAM	& 32GB & 128GB\\
				GPU	& NVIDIA RTX 3080TI & --\\
				Compiler& NVCC and GCC 9.4.0 & GCC 9.4.0\\
				OS	 &  Ubuntu 20.04 & Ubuntu 20.04\\
				Toolkit	& CUDA 12.1& --\\
				\bottomrule
			\end{tabular*}
		\end{minipage}
	\end{center}
\end{table}
\begin{table}[htbp]
	\begin{center}
		\begin{minipage}{\linewidth}
			\caption{Parameters of the Algorithms}\label{Algorithms}%
			\begin{tabular*}{\textwidth}{@{\extracolsep{\fill}}cl@{\extracolsep{\fill}}}
				\toprule
				Abbreviation &	Solution\\
				\midrule
				Gunrock	            & The BFS algorithm running on RXT3080TI,  \\
				& from the Gunrock \cite{Wang:2017:GGG,Osama:2022:EOP}\\
				GAP &  Direction-optimizing BFS algorithm, from \\
				&the GAP \cite{beamer2012direction}, running on I5-13600KF\\
				DAWN	        & DAWN with RTX 3080TI\\
				DAWN(20)	        & DAWN with I5-13600KF \\
				\bottomrule
			\end{tabular*}
		\end{minipage}
	\end{center}
\end{table}
\begin{table}[htbp]
	\begin{center}
		\caption{Parameters of the Experimental Graphs (Nodes and Edges)}\label{ExperimentalGraphs}
		\begin{minipage}{\linewidth}
			\begin{tabular*}{\textwidth}{@{\extracolsep{\fill}}ccccc@{\extracolsep{\fill}}}
				\toprule
				\multicolumn{5}{c}{\textbf{\large Nodes}}\\
				\midrule
				< 100K & 100K $\thicksim$ 500K & 500K $\thicksim$ 5M & 5M $\thicksim$ 100M& > 100M \\
				\midrule
				14 & 24 & 14&13& 1 \\
				\bottomrule
			\end{tabular*}
		\end{minipage}
		\begin{minipage}{\linewidth}
			\begin{tabular*}{\textwidth}{@{\extracolsep{\fill}}ccccc@{\extracolsep{\fill}}}
				\toprule
				\multicolumn{5}{c}{\textbf{\large Edges}}\\
				\midrule
				< 1M & 1M $\thicksim$ 5M & 5M $\thicksim$ 20M & 20M $\thicksim$ 500M & > 500M \\
				\midrule
				16 & 16 & 23 & 8 & 3 \\
				\bottomrule
			\end{tabular*}
		\end{minipage}
	\end{center}
\end{table}

The parameters of the test machine are detailed in Table \ref{TestMachine}. Our comparison includes various versions of the BFS algorithm, and the results are presented in Table \ref{Algorithms}. We provide accessible links to graphs: \href{https://www.scidb.cn/s/6BjM3a}{[Dataset]}. The number of nodes in these graphs ranges up to \textbf{$139M$}, with edges extending up to \textbf{$921M$}. The parameters of the experimental graphs are detailed in Table \ref{ExperimentalGraphs}.

Specifically, the results for DAWN running on GPUs were obtained using a thread block size of 1024, a configuration viable on GPUs since the Pascal architecture introduced in 2016. Although the optimal block partitioning scheme depends on several factors (e.g., matrix density, shared memory size, bandwidth, etc.), we adopt a fixed block size to enhance result reproducibility.

Gunrock is a CUDA library for graph-processing designed specifically for the GPU, which achieves a better balance between performance and expressiveness via coupling high-performance GPU computing primitives and optimization strategies, particularly in the area of fine-grained load balancing. \cite{Wang:2017:GGG,Osama:2022:EOP,10.1145/3572848.3577434,10.1145/2851141.2851145}.

We strongly encourage readers to delve into the provided codebase and verify the reported results. The code and more information for our algorithm are available on \href{https://github.com/lxrzlyr/DAWN-An-Noval-SSSP-APSP-Algorithm}{[GitHub]}. In the repository, we offer additional insights into the actual running times and graph details for each proposed solution, accompanied by a description of artifacts and evaluation methodologies. These details are provided to enhance the reproducibility of any results presented in this paper.

\subsection{Scalability}\label{scalability}
It is important to validate DAWN's feature of high parallelism and scalability. We measure the scalability of DAWN using multi-threading efficiency as follows, simplified from Gustafson-Barsis's law\cite{gustafson1988reevaluating},
\begin{equation}
	\eta_t=\frac{T_B}{T_N \times N},
\end{equation}
where $T_B$ represents the baseline execution time, $T_N$ represents the execution time of the program with $N$ threads, and $N$ is the number of threads. In Table \ref{tab:CPUI5} and \ref{tab:CPUAMD}, the multi-threading efficiency for DAWN based on SOVM and BFS API from GAP on the I5-13600KF and EPYC Milan 7T83 are depicted, respectively.

\begin{table}[htbp]
	\caption{The multi-threading efficiency of DAWN and GAP on I5-13600KF}\label{tab:CPUI5}
	\begin{tabular}{|c|c|c|c|c|c|}
		\toprule
		Thread   & 1     & 3       & 6       & 12      & 20      \\
		\midrule
		DAWN(20) & 100\% & 99.72\% & 98.35\% & 77.96\% & 37.54\% \\
		GAP      & 100\% & 96.84\% & 89.51\% & 66.28\% & 23.43\% \\
		\bottomrule
	\end{tabular}
\end{table}

\begin{table}[htbp]
	\caption{The multi-threading efficiency of DAWN and GAP on EPYC Milan 7T83}\label{tab:CPUAMD}
	\begin{tabular}{|c|c|c|c|c|c|}
		\toprule
		Thread   & 4     & 8       & 16      & 32      & 64      \\
		\midrule
		DAWN(20) & 100\% & 99.45\% & 97.73\% & 84.10\% & 58.97\% \\
		GAP      & 100\% & 95.69\% & 88.12\% & 71.49\% & 40.16\% \\
		\bottomrule
	\end{tabular}
\end{table}

When utilizing up to 32 threads on the EPYC processor, the core frequency remains constant at 3.5 GHz. However, when scaling up to 64 threads, the frequency of all cores diminishes to 2.54 GHz. In contrast, the I5 processor does not experience any reduction in core frequency. It is important to note that the exact performance gains are contingent upon the particular hardware configuration utilized, and considerations such as power and thermal constraints impose limitations on the maximum achievable performance.

The I5 processor integrates a combination of performance and efficiency cores, where the former delivers higher clock speeds, and the latter excels in power efficiency. Hence, DAWN achieves a linear speedup when scaling from 1 thread to 6 threads, but the performance improvement slows down when scaling from 6 to 12 threads due to the performance gap between the two types of cores. Additionally, the I5 processor does not have enough physical cores to achieve significant performance gains beyond 14 threads.

Across diverse hardware configurations, DAWN demonstrates better multi-threading efficiency compared to GAP. Futhurmore, the scalability of the algorithm is influenced by a considerable number of factors, among which the characteristic of the graph is a significant factor.

\begin{figure}[htbp]
	\centering
	\includegraphics[width=\linewidth]{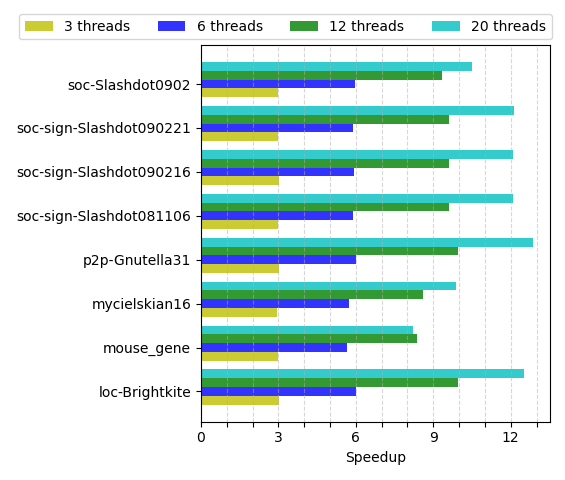}
	\caption{Speedup for DAWN based on SOVM in various threads (baseline 1 thread), with Intel Core i5-13600KF baseline of 1 thread}
	\label{fig:scalability20}
\end{figure}

\begin{figure}[htbp]
	\centering
	\includegraphics[width=\linewidth]{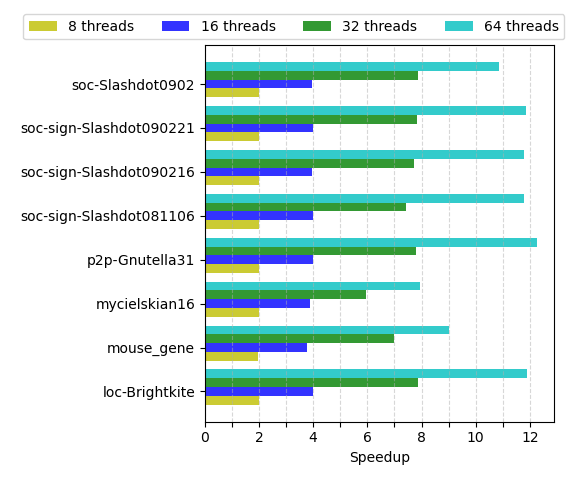}
	\caption{Speedup for DAWN based on SOVM in various threads (baseline 4 threads), with AMD EPYC Milan 7T83 }
	\label{fig:scalability64}
\end{figure}

Figure \ref{fig:scalability20} and \ref{fig:scalability64} illustrate DAWN's capability for speedup across different thread counts on some of the graphs. Specifically, on the \textit{mycielskian16}, which is a dense graph with a low-diameter of 2, DAWN exhibits lower thread efficiency compared to other sparse graphs. This phenomenon underscores the impact of multiple factors on the increase or decrease in algorithm performance. For instance, Graph \textit{mouse\_gene} is denser than \textit{mycielskian16}, and with a diameter of 12, yet DAWN exhibits superior thread efficiency on \textit{mouse\_gene}. Therefore, we emphasize the comprehensive performance of the algorithm across a wider variety of graphs.
\subsection{Performance Comparison with GAP}

In the experiment, DAWN based on SOVM demonstrated remarkable performance and also exhibits high scalability and parallelism. We have included a figure illustrating the distribution of the speedup of DAWN over GAP across all graphs in the test dataset.

\begin{table}[htbp]
	\caption{The speedup of DAWN over GAP}\label{tab:CPUspeedup}
	\begin{tabular}{|c|c|c|c|c|c|}
		\toprule
		Speedup  & <1$\times$ & 1$\times$ $\thicksim$ 2$\times$ & 2$\times$ $\thicksim$ 4$\times$ & 4$\times$ $\thicksim$ 16$\times$ & >16$\times$ \\
		\midrule
		DAWN(20) & 4          & 15                              & 24                              & 17                               & 6           \\
		DAWN     & 12         & 20                              & 11                              & 9                                & 14          \\
		\bottomrule
	\end{tabular}
\end{table}

\begin{figure*}[htbp]
	\centering
	\includegraphics[width=\textwidth]{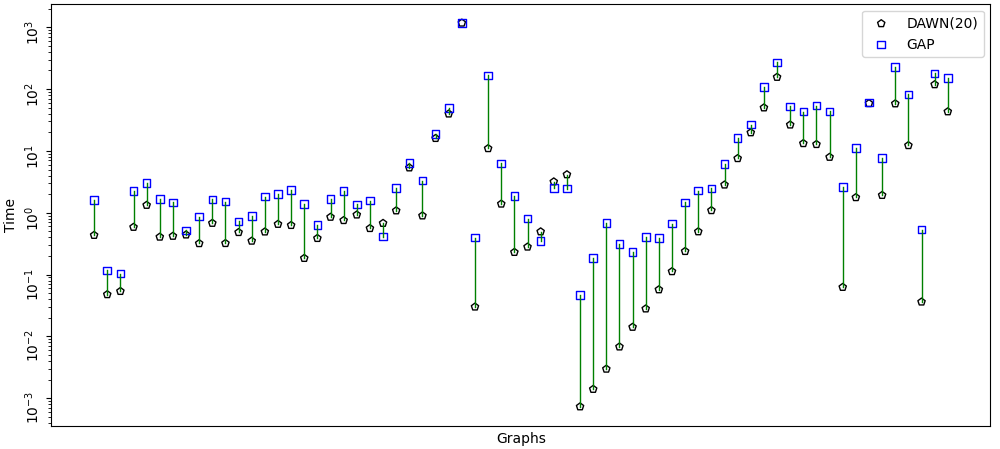}
	\caption{Running time for the DAWN based on the SOVM and BFS API from GAP with an I5-13600KF}
	\label{fig:CPU}
\end{figure*}
\begin{figure*}[htbp]
	\centering
	\includegraphics[width=\textwidth]{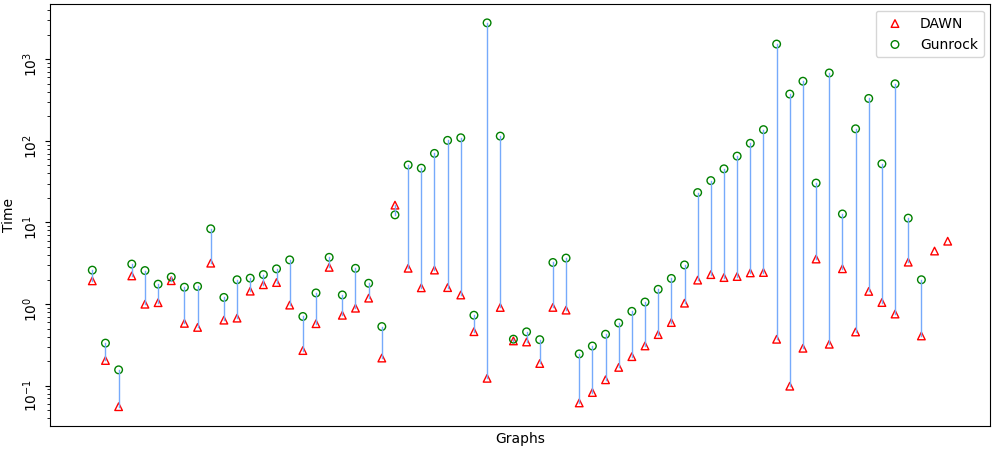}
	\caption{Running time for the DAWN based on the SOVM and BFS API from Gunrock with an RXT3080TI}
	\label{fig:GPU}
\end{figure*}

In Table \ref{tab:CPUspeedup}, the speedup for DAWN based on SOVM over BFS API from GAP on an I5-13600KF is depicted, with the values derived from the mean of repeated experiments, following the methodology outlined in Section \ref{exp}. The first row shows the speedup of DAWN(20) over BFS API from GAP, both on I5-13600KF, and the next row shows the speedup of DAWN on RTX3080TI over BFS API from GAP on I5-13600KF. Due to the significant increase in scalability and parallelism, DAWN based on the SOVM outperformed GAP in most graphs (62 out of 66), achieving an impressive average speedup of \textbf{3.769$\times$}.

However, the DAWN algorithm demonstrates comparatively lower performance in four specific graphs (\textit{coPapersDBLP}, \textit{com-DBLP}, \textit{coAuthorsDBLP}, \textit{coPapersCiteseer}), all representing citation and collaboration networks. These graph types are characterized by high clustering coefficients and relatively short average shortest paths. Despite the deployment of a more potent processor, BFS API from Gunrock falls short of surpassing the performance of GAP on these graphs. Nevertheless, in other scale-free graphs such as social networks and the internet, the DAWN algorithm exhibits superior performance.

Numerous well-established studies have presented evidence that the eccentricity of the real graphs is $\log n$ \cite{bollobas1981diameter,albert1999diameter, PhysRevLett.90.058701,riordan2004diameter,ganesh_xue_2007,10.1145/2063576.2063748}. Therefore, we get the small-world graphs (23 out of 66) which the average shortest path in the graph is less than $\log n$, includes the citation and collaboration networks mentioned before.

The Direction-Optimizing BFS algorithm will achieve the speedups when the active frontier is a substantial fraction of the total graph, which commonly occurs in small-world graphs \cite{beamer2012direction}. However, DAWN outperforms GAP on the most small-world graphs (19 out of 23) and achieves an average speedup of  \textbf{2.332$\times$}. Furthermore, in other real graph with a high-diameter such as road networks, DAWN achieves an average speedup of \textbf{4.483$\times$} over GAP.

Figure \ref{fig:CPU} shows the running time for DAWN(20) and GAP. The y-axis represents the average running time, with each marker representing the running time on a graph. GAP instances are indicated by blue markers, while DAWN instances are represented by black markers.

\subsection{Performance Comparison with Gunrock}
In Table \ref{tab:GPUspeedup}, the first row shows the speedup of DAWN(20) on an I5-13600KF over BFS API from Gunrock on RTX3080TI. The next rows shows the speedup of DAWN over BFS API from Gunrock, both on RTX3080TI. Figure \ref{fig:GPU} illustrates the running time for DAWN and BFS API from Gunrock. Red markers correspond to DAWN, while green markers represent Gunrock. Impressively, DAWN outperformed Gunrock in the majority of graphs (63 out of 66), achieving an average speedup of \textbf{9.410$\times$}. On the Graphs \textit{uk-2005} and \textit{arabic-2005}, Gunrock encountered an out-of-memory error, thus preventing the acquisition of runtime data for these two graphs. The testing machine equipped with 12GB of physical GPU memory, with 9.7GB available. The available GPU memory for both DAWN and Gunrock is identical, indicating that when executing similar tasks, DAWN requires less GPU memory compared to Gunrock.

\begin{table}[htbp]
	\caption{The speedup of DAWN over Gunrock}\label{tab:GPUspeedup}
	\begin{tabular}{|c|c|c|c|c|c|}
		\toprule
		Speedup  & <1$\times$ & 1$\times$ $\thicksim$ 2$\times$ & 2$\times$ $\thicksim$ 4$\times$ & 4$\times$ $\thicksim$ 16$\times$ & >16$\times$ \\
		\midrule
		DAWN(20) & 5          & 5                               & 10                              & 23                               & 21          \\
		DAWN     & 3          & 16                              & 22                              & 6                                & 19          \\
		\bottomrule
	\end{tabular}
\end{table}

Apart from the aforementioned two graphs, DAWN demonstrates performance inferior to that of Gunrock on the Graph \textit{web-BerkStan}, and also falls short compared to DAWN(20) and GAP running on CPU. This phenomenon may be attributed to the scale-free nature of {web-BerkStan}, leading to load imbalance during computation and significantly impacting algorithm performance. Gunrock, possessing robust load balancing capabilities, holds an advantage in such scenarios. It is crucial to note that DAWN does not prioritize load balancing as the primary focus of our investigation. Nonetheless, despite these challenges, DAWN outperforms Gunrock on the majority of graphs due to algorithmic optimizations.

\subsection{Performance on Different Platforms}

\begin{figure}[htbp]
	\centering
	\includegraphics[width=\linewidth]{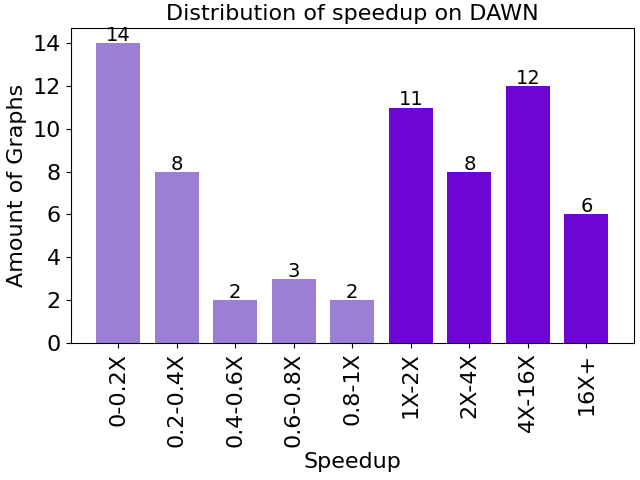}
	\caption{The speedup of DAWN(20) over DAWN}
	\label{fig:dawnspeedup}
\end{figure}

The differences in the speedup distribution of DAWN compared to different algorithms are attributed to the nature of the graphs, such as the number of nodes and the average shortest path length. Therefore, we will proceed to compare the performance of DAWN on the different platforms. Figure \ref{fig:dawnspeedup} illustrates the performance gap of DAWN between CPU and GPU. Light purple bars indicate cases where DAWN's performance on CPU is inferior to that on GPU, while dark purple bars represent the opposite scenario.

In more than half of the graphs (37 out of 66), DAWN(20) exhibits superior performance compared to DAWN. For instance, on web graphs, DAWN(20) and the GAP algorithm achieved enhanced performance, which demonstrates that the powerful single-core performance of CPUs provide better acceleration for algorithms.

However, this single-core performance acceleration has its limitations. Once the graph size exceeds one million nodes, the advantage of single-core performance can no longer compensate for the performance disparity induced by a greater number of cores. Furthermore, on graphs with a smaller number of nodes, the communication overhead between the CPU and GPU appears more costly than computational expenses, leading to inferior performance compared to algorithms running on CPU.

DAWN(20) achieved performance superiority on graphs with an average of 0.209 million nodes and 5.854 million edges (considering undirected edges as two directed edges). On the other hand, DAWN demonstrated performance superiority on graphs with an average of 13.820 million nodes and 146.592 million edges.

In summary, DAWN is more efficient and yielding a higher speedup when compared to Gunrock and GAP.

\section{Conclusion}
In this paper, we propose an enhanced BFS algorithm, which requires $O(E_{wcc}(i))$ and $O(S_{wcc} \cdot E_{wcc})$ time for solving SSSP and APSP problems on the unweighted graph, respectively.

Our research involved a performance evaluation of DAWN, GAP, and Gunrock across various platforms, using graphs from SuiteSparse Matrix Collection and Gunrock benchmark dataset. DAWN achieves average speedup of 3.769$\times$ and 9.410$\times$, over GAP and Gunrock respectively.

The experiment underscores that DAWN based on the SOVM, exhibits remarkable scalability and efficiency in addressing the shortest paths problem on modern processors. The efficient utilization of computational resources is a significant factor contributing to its exceptional performance. These results highlight the potential of DAWN as a powerful tool for graph analytics, particularly in applications that require high processing speed and efficiency.

Our future research will focus on addressing the balance between optimizing matrix operations and managing the consumption of (min,+) operations. This focus is aimed to expand the applicability of DAWN on weighted graphs, transforming it from a promising proof-of-concept to a practical tool that can be used in various real-world graph analysis applications in the future.

\bibliographystyle{ACM-Reference-Format}
\bibliography{bibfile}

\end{document}